\documentclass[aps,prl,twocolumn,superscriptaddress,showpacs,showkeys]{revtex4}

\usepackage{graphicx}

\begin{document}


\title{Light Scattering by Low Lying Quasiparticle Excitations in the Fractional Quantum Hall Regime}


\author{C.F. Hirjibehedin}
\affiliation{Department of Physics, Columbia University, New York, NY
10027} \affiliation{Bell Labs, Lucent Technologies, Murray Hill, NJ
07974}

\author{Irene Dujovne}
\affiliation{Department of Appl. Physics and Appl. Mathematics,
Columbia University, New York, NY  10027} \affiliation{Bell Labs,
Lucent Technologies, Murray Hill, NJ 07974}

\author{A. Pinczuk}
\affiliation{Department of Physics, Columbia University, New York, NY
10027} \affiliation{Bell Labs, Lucent Technologies, Murray Hill, NJ
07974} \affiliation{Department of Appl. Physics and Appl.
Mathematics, Columbia University, New York, NY  10027}

\author{B.S. Dennis}
\affiliation{Bell Labs, Lucent Technologies, Murray Hill, NJ 07974}

\author{L.N. Pfeiffer}
\affiliation{Bell Labs, Lucent Technologies, Murray Hill, NJ 07974}

\author{K.W. West}
\affiliation{Bell Labs, Lucent Technologies, Murray Hill, NJ 07974}

\date{\today}

\begin{abstract}
Low lying excitations of electron liquids in the fractional quantum
Hall (FQH) regime are studied by resonant inelastic light scattering
methods. We present here results from charge and spin excitations of
FQH states in the lowest spin-split Landau levels that are of current
interest. In the range of filling factors $2/5 \geq \nu \geq 1/3$, we
find evidence that low energy quasiparticle excitations can be
interpreted with spin-split composite fermion quasi-Landau levels. At
FQH states around $\nu=3/2$, we find well-defined excitations at
$4/3$ and $8/5$ that are consistent with a spin-unpolarized
population of quasi-Landau levels.
\end{abstract}



\maketitle


The fractional quantum Hall effect (FQHE) occurs at low temperatures
in two dimensional (2D) electron systems of very low disorder that
are embedded in large perpendicular magnetic fields. Condensation of
2D electron systems into quantum fluids is among the striking
manifestations of fundamental interactions in two dimensions. The
major sequences of the FQHE occur at electron Landau level (LL)
filling factors given by $\nu = p / (\phi p \pm 1)$, where $p$ is an
integer that enumerates members of a particular sequence and $\phi$
is an even integer that labels different sequences.
\par

\begin{figure}
\includegraphics{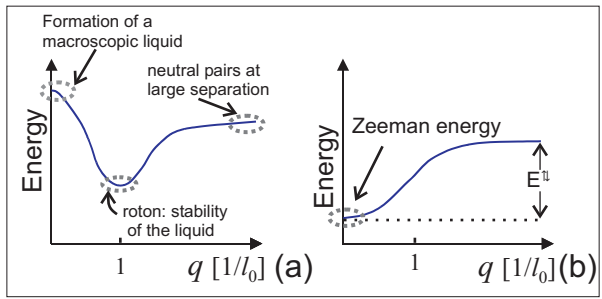}
\caption{\label{fig:disp} Schematic representation of the wavevector
dispersions of low lying excitations in the FQHE regime. (a) Charge
modes. (b) Spin wave modes. The wavevector is represented in units of
$1/l_0$.}
\end{figure}

Excitations of quasiparticles above the FQHE ground state are
described as modes of energy $\omega$ that have wavevector
dispersions $\omega (q)$. Fig. \ref{fig:disp} shows schematic
renditions of low lying quasiparticle excitations at $\nu=1/3$. Panel
(a) is for the charge modes in which there are no changes in spin
orientation. The mode at large wavevector ($q\rightarrow\infty$),
which represents a neutral quasiparticle-quasihole pair at large
separation \cite{Kallin&Halperin1984}, is understood as the energy
obtained in activated magnetotransport. The other salient features of
the mode dispersions are the $q \rightarrow 0$ limit and the minima
in the dispersion at finite wavevector $q \sim 1/l_0$, where $l_0 =
(\hbar c / e B)^{1/2}$ is the magnetic length. The minima, called
magnetorotons or simply rotons, are the consequences of excitonic
binding interactions between the quasiparticles
\cite{Kallin&Halperin1984,Haldane&Rezayi1985,Girvin1985}. Panel (b)
of Fig. \ref{fig:disp} describes the dispersion of spin wave
excitations in which the only change is the orientation of
quasiparticle spin
\cite{Longo&Kallin1993,Nakajima&Aoki1994,Chakraborty2000,Mandal2001a}.
Here the $q \rightarrow 0$ limit is at the bare Zeeman energy E$_{z}$
and the large wavevector limit ($q\rightarrow\infty$) is shifted by
the interaction energy required to reverse a spin orientation
E$^{\uparrow\downarrow}$.
\par

\begin{figure}
\includegraphics{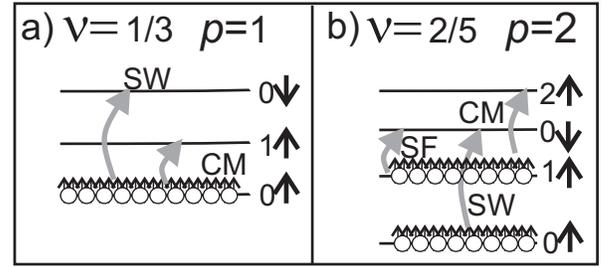}
\caption{\label{fig:levels} Schematic representation of lowest
spin-split energy levels of CF quasiparticles. Numbers and black
arrows indicate the CF-LL index and spin orientation. (a) $\nu=1/3$,
$p=1$. (b) $\nu=2/5$, $p=2$. Grey arrows represent the transitions in
low lying collective excitations. In CM transitions there is only
change in level index. In SW transitions there is only change in spin
orientation. In SF transitions level index and spin orientation
change.}
\end{figure}

In the composite fermion (CF) picture $\phi$ vortices of the
many-body wavefunction are attached to each electron to construct CF
quasiparticles \cite{Jain1989,Heinonen}. Within this framework vortex
attachment takes into account Coulomb interactions and $p$ is the CF
filling factor. CF quasiparticles have spin-split LLs characteristic
of charged fermions with spin 1/2. Chern-Simons gauge fields
incorporate electron interactions so that CFs in the lowest electron
LL ($\nu \leq 1$) experience effective magnetic fields $B^*= B -
B_{1/\phi} = \pm B/(\phi p \pm 1)$, where $B$ is the perpendicular
component of the external field
\cite{Lopez&Fradkin1991,Kalmeyer&Zhang1992,HalperinLeeRead}.
Spin-split CF-LLs are shown schematically in Fig. \ref{fig:levels}
for $\nu=1/3$ in panel (a) and $\nu=2/5$ in panel (b). The spacing
between lowest CF levels with same spin is described as a cyclotron
frequency
\cite{Lopez&Fradkin1991,HalperinLeeRead,Lopez&Fradkin1993,Du1993,Park1998a,Murthy1999,Mandal2001a,Onoda2001}
\begin{equation}
  \omega_{CF}=\frac{e B^*}{c m_{CF}}
  \label{wcf}
\end{equation}
and $m_{CF}$ is a CF effective mass. In the context of dispersive
collective excitations the large wavevector limit
($q\rightarrow\infty$) of the lowest charge excitations is at energy
$\omega_{CF}$.
\par

The main sequence of the FQHE corresponds to $\phi=2$. In this
sequence there are $p$ fully occupied CF-LLs, as shown for $p=1$ and
$p=2$ in Fig. \ref{fig:levels}. The quasiparticle excitations shown
in Fig. \ref{fig:disp} are constructed from the neutral
quasiparticle-quasihole pairs of the transitions shown in panel (a)
of Fig. \ref{fig:levels}. The CM transitions are for charge modes and
the SW transitions for spin wave excitations.
\par

Given that vortex attachment incorporates effects of Coulomb
interactions through Chern-Simons gauge fields, CFs have residual
interactions that are much weaker than those among electrons.
Features of wavevector dispersions of CF excitations such as rotons
and many-body spin reversal energies can be regarded as
manifestations of residual interactions among CF quasiparticles.
\par

The low lying excitations of CF quasiparticles, with typical energies
below $1 \mathrm{meV}$, are being studied by resonant inelastic light
scattering methods
\cite{Kang2001,Dujovne2003a,Dujovne2003b,Hirjibehedin2003a}. In this
paper we consider recent light scattering studies of low lying
quasiparticle excitations at filling factors in the range $2/5 \geq
\nu \geq 1/3$ \cite{Dujovne2003a,Dujovne2003b}. We also consider
results obtained in FQHE liquids in higher electron LLs at filling
factors in the range $2 > \nu > 1$.
\par

The main features of dispersions of low lying quasiparticle
excitations of spin and charge collective modes in the FQHE were
reported in resonant inelastic light scattering experiments
\cite{Pinczuk1993,Davies1997,Kang2000}. More recent light scattering
results at filling factors $1/3$ and $2/5$ have determined the
energies of rotons at $q \sim 1/l_0$ and of large wavevector
($q\rightarrow\infty$) excitations of CFs
\cite{Kang2001,Dujovne2003a,Dujovne2003b}. The light scattering
results for spin and charge modes, quantitatively explained within CF
theory
\cite{Kang2001,Scarola2000a,Mandal2001a,Dujovne2003a,Dujovne2003b},
suggest that the structure of CF-LLs and residual interactions can be
explored by light scattering measurements of low lying collective
excitations. The residual interactions that manifest in light
scattering experiments considered here could lead to formation of
higher order quantum liquids such as those probed in recent
magnetotransport experiments \cite{Pan2003}.
\par

\section{Experimental Setup and Methodology}
\label{sec:Setup}

\begin{figure}
\includegraphics{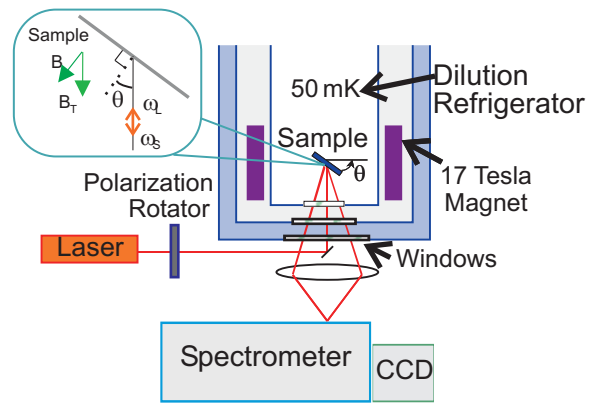}
\caption{\label{fig:setup} A schematic of the experimental setup and
measurement configuration. The inset shows the backscattering
geometry.}
\end{figure}

The high quality 2D electron system studied here is formed in an
asymmetrically doped, $330\mathrm{\AA}$-wide GaAs single quantum well
(SQW). The sample density is $n = 5.6\times 10^{10} \mathrm{cm}^{-2}$
and the electron mobility is $\mu \gtrsim 7 \times 10^{6}
\mathrm{cm}^{2}/\mathrm{Vsec}$ at $T = 300 \mathrm{mK}$. As shown in
Fig. \ref{fig:setup}, samples are mounted on the cold finger of a
dilution refrigerator with a base temperature of $50\mathrm{mK}$ that
is inserted into the cold bore of a $17 \mathrm{T}$ superconducting
magnet. The samples are mounted with the normal to the surface at an
angle $\theta$ to the magnetic field $B_T$, making the component of
the field perpendicular to the 2D electron layer $B = B_T cos
\theta$.
\par

As shown in the schematic in Fig. \ref{fig:setup}, light scattering
measurements are performed through windows for direct optical access
to the sample. The energy of the linearly polarized incident photons
$\omega_L$ is tuned close to the fundamental optical transitions of
the GaAs well with either a diode or dye laser. The power density is
kept below $10^{-4} \mathrm{W}/\mathrm{cm}^2$. The samples are
measured in a backscattering geometry, making an angle $\theta$
between the incident/scattered photons and the normal to the sample
surface. The wavevector transferred from the photons to the 2D system
is $k=k_L^{||} - k_S^{||} = (2 \omega_L / c) sin \theta$, where
$k_L^{||}$ and $k_S^{||}$ are the components of the incident and
scattered photon wavevectors parallel to the 2D plane and the
difference between $\omega_L$ and the scattered photon energy
$\omega_S$ is small. For typical measurement geometries, $\theta <
60^{\circ}$ so that $k \leq 1.5 \times 10^5 cm^{-1} \ll 1/l_0$.
Conservation of energy in the inelastic scattering processes is
expressed as $\omega(q)=\pm (\omega_L-\omega_S)$, where $q$ is the
wavevector of the excitation and $+$ ($-$) corresponds to the Stokes
(anti-Stokes) process.
\par

Scattered light is dispersed by a Spex 1404 double Czerny-Turner
spectrometer operating in additive mode with holographic master
gratings that reduce the stray light. Photons are detected with CCDs
with $15$ or $20 \mu \mathrm{m}$ pixels.  The combined resolution of
the system with entrance slits on the spectrometer set to $30 \mu
\mathrm{m}$ is $16 \mu \mathrm{eV}$. The response of the spectrometer
is linearly polarized, so that spectra can be taken with the linear
polarization of the incident photons parallel (polarized) or
perpendicular (depolarized) to the detected scattered photons'
polarization. Excitation modes with changes in the spin degree of
freedom tend to be stronger in depolarized spectra, while modes in
the charge degree of freedom tend to be stronger in polarized
spectra.
\par

\section{Excitations at $\nu=1/3$}
\label{sec:1/3}

\begin{figure}
\includegraphics{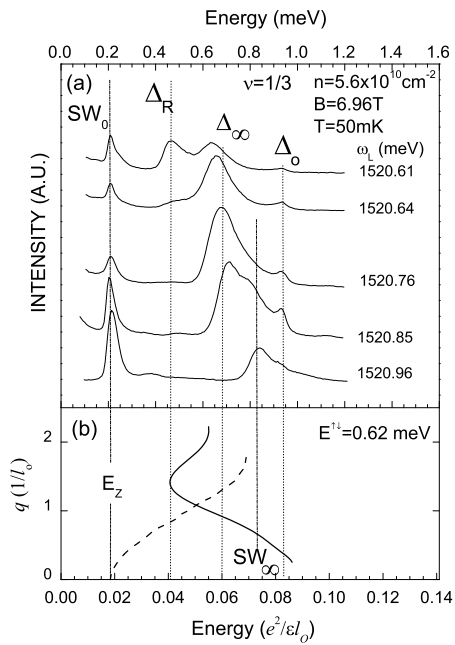}
\caption{\label{fig:1/3}(a) Resonant inelastic light scattering
spectra at $\nu=1/3$ for different $\omega_L$. (b) Dispersion curves
for the charge mode and the spin wave. The $SW_0$ excitation is found
at the bare Zeeman energy (Larmor's theorem)
\cite{Kallin&Halperin1984} E$_z$. The value of
E$^{\uparrow\downarrow}$ is obtained from Refs.
\cite{Dujovne2003a,Dujovne2003b}.}
\end{figure}

Low energy excitations at $\nu=1/3$ include charge modes, where the
CF-LL quantum number changes and spin modes where the spin degree of
freedom is modified. At $\nu=1/3$ only the lowest CF-LL is populated
and the lowest energy excitations correspond to $|0\uparrow>
\rightarrow |1\uparrow>$ and $|0\uparrow> \rightarrow |0\downarrow>$
CF-LL transitions for the CM and SW respectively, as shown in Fig.
\ref{fig:levels}.
\par

Figure \ref{fig:1/3}(a) shows resonant inelastic light scattering
spectra at $\nu=1/3$ for different values of $\omega_L$. When
$\omega_L$ is close to the fundamental optical transitions of the 2D
system, the intensity of the inelastic light scattering is increased
dramatically
\cite{Burstein1979,Burstein1980,Pinczuk1988,Pinczuk1993,Pinczuk1994,Platzman1994,Davies1997,DasSarma1999,Hirjibehedin2003b}.
As seen in Fig. \ref{fig:1/3}(a), the various excitations observed at
$1/3$ have different resonance conditions. For many of these modes,
resonances occur when $\omega_L$ or $\omega_S$ overlap transitions
seen in photoluminescence \cite{Hirjibehedin2003b} that are assigned
to recombination of negatively charged excitons
\cite{Kheng1993,Shields1995,Wojs2000,Yusa2001}. A significant feature
of the spectra in Fig. \ref{fig:1/3}(a) is the absence of
luminescence. This is characteristic of spectra excited with
$\omega_L$ that overlap the lowest optical excitation of the GaAs
quantum well \cite {Hirjibehedin2003b}.
\par

The dispersion curves for the CM \cite{Scarola2000a} and the SW
\cite{Nakajima&Aoki1994} excitations are displayed in Fig.
\ref{fig:1/3}(b). The calculations have been scaled by a factor of
$0.6$ to account for the effects of the finite width of the 2D layer
\cite{Zhang&DasSarma1986,Park1998b,Kang2001,Hirjibehedin2003a}. The
main features of the dispersions are at critical points that occur at
$q \rightarrow 0$, at the $q \rightarrow \infty$ limit, and at the
roton minimum at $q \sim 1/l_0 \sim 10^{6}cm^{-1}$. It is clear that
the spectra in panel (a) display the features of the critical points
of the wavevector dispersions of the excitation modes that have
wavevectors $q>>k\sim 10^{5}cm^{-1}$. The results in Fig.
\ref{fig:1/3} indicate a massive breakdown of wavevector conservation
in the resonant inelastic light scattering spectra.
\par

In a translationally invariant system conservation of momentum is
equivalent to conservation of wavevector. Because the wavevector
transferred to the 2D system by the photons is $k \ll 1/l_0$, light
scattering in a translationally invariant system, where $q = k$, can
only access excitations with small wavevectors $q \ll 1/l_0$. The
original light scattering experiments in the FQHE regime were able to
access the $q \rightarrow 0$ low energy CM and SW excitations
\cite{Pinczuk1993}, labelled $\Delta_0$ and $SW_0$ respectively in
Fig. \ref{fig:1/3}(a).
\par

Breakdown of wavevector conservation allows inelastic light
scattering to access the energies of rotons and of large wavevector
($q\rightarrow\infty$) excitations of the CM and SW excitations
\cite{Kang2001,Dujovne2003b}. The peaks labelled $\Delta_R$ and
$\Delta_{\infty}$ in Fig. \ref{fig:1/3}(a) have been assigned to the
roton and to $q \rightarrow \infty$ CM excitations
\cite{Davies1997,Pinczuk1998,Kang2001}. The mode labelled
$SW_{\infty}$ has been assigned to the $q \rightarrow \infty$ spin
wave excitation \cite{Dujovne2003b}. The $SW_{\infty}$ energy is
E$_z^* = \mathrm{E}_{z} +\mathrm{E}^{\uparrow\downarrow}$ where
E$^{\uparrow\downarrow}$ represents the spin reversal energy due to
interactions among quasiparticles
\cite{Chakraborty2000,Longo&Kallin1993,Nakajima&Aoki1994,Lopez&Fradkin1995,Mandal2001b,Dujovne2003a,Dujovne2003b}.
The value of E$^{\uparrow\downarrow}=0.054 e^2/ \epsilon l_0$
obtained from the $SW_{\infty}$ energy is consistent with that
obtained from the energies of spin flip excitations ($|1\uparrow>
\rightarrow |0\downarrow>$) for filling factors $\nu \gtrsim 1/3$
with small populations of the $|1\uparrow>$ CF-LL
\cite{Dujovne2003a,Dujovne2003b}.
\par

Breakdown of wavevector conservation in resonant inelastic light
scattering is attributed to the loss of full translation symmetry in
the presence of very weak residual disorder. The breakdown is
particularly effective for spectra excited under strongly resonant
conditions. Here the higher order terms in a perturbation theory
description of light scattering that directly incorporate wavevector
relaxation processes are greatly enhanced \cite{Gogolin&Rashba1976}.
Quantum Hall systems of high perfection display extremely sharp
optical resonances of widths $< 0.2meV$ \cite{Hirjibehedin2003b}, as
in Fig. \ref{fig:1/3}. In such systems, light scattering processes
that break wavevector conservation can be extremely strong. For
strongly resonant conditions, light scattering can easily access
excitation modes with wavevectors far beyond $k$
\cite{Pinczuk1988,Davies1997,Pinczuk1998,Kang2001,Dujovne2003b}.
\par

Under conditions of conservation of wavevector, the intensity of
light scattering is proportional to the dynamic structure factor
$S(q=k,\omega)$ \cite{Abstreiter1984}. Phenomenologically, the light
scattering intensity with breakdown of wavevector conservation can be
represented by a dynamical structure factor defined as
\cite{Marmorkos&DasSarma1992}
\begin{equation}
S(k,\omega; \alpha) \sim \int d Q f(k,Q; \alpha) S(Q,\omega)
\end{equation}
where $\alpha$ is a parameter (or set of parameters) associated with
the diverse mechanisms that contribute to breakdown of wavevector
conservation and $f(k,Q; \alpha)$ is a function that couples
structure factors at different wavevectors. Within this framework,
$S(k,\omega;\alpha)$ and therefore the intensity of the scattered
light contains contributions from modes with wavevectors which are
well beyond the $q = k$ limit.
\par

Given the importance of breakdown of wavevector conservation in
resonant inelastic light scattering processes in the integer and
fractional quantum Hall regimes it is useful to conjecture about
possible mechanisms that contribute to $\alpha$. We may envision two
classes of processes. One class arises from processes that contribute
to inhomogeneous broadening and lifetimes of optical transitions of
the GaAs quantum wells. Fluctuations in quantum well widths typically
contribute to such processes. Another important class of
contributions is that arising from wavevector relaxation processes
responsible for the increase in magnetoresistance that signals the
onset of dissipation in the quantum liquid with increasing
temperature.

\section{States with $2/5 \geq \nu \geq 1/3$ and $\nu \leq 1/3$}
\label{sec:between}

The main FQH states linked to CFs that bind two vortices ($\phi=2$)
form the sequence $\nu=p/(2p\pm1)$, where, as mentioned above, $p$
represents the CF filling factor and is an integer at the main FQHE
states. As shown in Fig. \ref{fig:levels}, at $\nu=1/3$ there is only
one CF-LL populated while at $\nu=2/5$ the lowest two levels are
full. Inelastic light scattering measurements of low lying
excitations of the electron liquids in the full range of filling
factors $2/5 \geq \nu \geq 1/3$ enable the study of quasiparticles in
states with partial population of one CF-LL ($2 \geq p \geq 1$).
These experiments uncover low lying excitations at all magnetic
fields within this range \cite{Dujovne2003a}. There has been recent
interest in the region between major fractions such as $2/5 > \nu >
1/3$ where newly observed FQH states occur at values such as
$p=1+1/3$ and $p=1+2/3$
\cite{Pan2003,Szlufarska2001b,Scarola2002,Chang2003}. Such liquids
may be described by attaching an even number of fluxes to CFs,
resulting in quasiparticles with $\phi=4$.
\par

\begin{figure}
\includegraphics{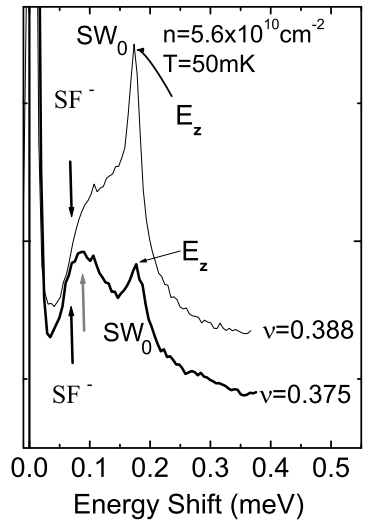}
\caption{\label{fig:roton} Depolarized inelastic light scattering
spectra at filling factors $\nu=$ 0.388 (thin line) and 0.375 (thick
line) near $\nu=2/5$. Vertical solid arrows show the positions of
roton density of states as presented in Ref. \cite{Dujovne2003a}. The
gray arrow marks the position of the maximum of the peak at
$\nu=0.375$.}
\end{figure}

Figure \ref{fig:roton} shows light scattering spectra of low lying
excitations ($\omega \lesssim 0.3\mathrm{meV}$) measured at $\nu
\lesssim 2/5$. The spectrum at $\nu=0.388$ shows two structures. The
one labelled (SW$_0$) is at E$_z=g\mu_B B_T$. At lower energies,
there is a broader structure labelled SF$^-$ that is assigned to an
excitation based on transitions similar to SF transitions shown in
Fig. \ref{fig:levels}(b) \cite{Dujovne2003a,Dujovne2003b}. The mode
involves simultaneous changes in the CF-LL quantum number and spin
orientation \cite{Mandal2001a}. In Refs.
\cite{Dujovne2003a,Dujovne2003b} it is shown that SF$^-$ evolves from
a deep roton minimum of SF excitations at $\nu=2/5$. As the filling
factor is reduced the SF$^-$ evolves into the well-defined peak shown
in Fig. \ref{fig:roton} for $\nu=0.375$. At larger magnetic fields
$\nu \gtrsim 1/3$ a narrow peak (SF$^+$) appears below the SW$_0$
(not shown in Fig. \ref{fig:roton}). This peak disappears at
$\nu=1/3$.
\par

\begin{figure}
\includegraphics{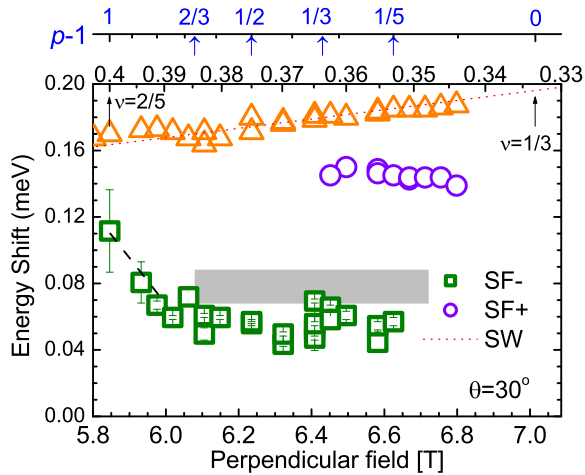}
\caption{\label{fig:energies} Energies of the low energy modes in the
filling factor range 2/5$\geq \nu \geq 1/3$ (after
\cite{Dujovne2003a}). The modes represented are the SW$_0$, SF$^+$,
and SF$^-$ described in the text. Two values are presented for SF$^-$
modes: the squares are determined from the roton density of states
\cite{Dujovne2003a}; the gray area covers the energy range of the
maxima of the SF$^-$ peak. The dotted line is a fit of the SW$_0$
energy with E$_{z} = g \mu_{B} B_T$ and $g = 0.44$.}
\end{figure}

Figure \ref{fig:energies} summarizes the energies of the low lying
spin excitations measured in the range $2/5 \geq \nu \geq 1/3$. The
mode energies shown include the SW$_{0}$ at E$_z$, the SF$^-$, and
the SF$^+$. The SF$^-$ mode displays rapid softening to very low
energies and then its energy remains constant until it disappears. It
is noteworthy that the energy of the SF$^-$ mode does not display a
marked magnetic field dependence in the region in which the filling
factor of the second CF-LL, given by $p-1$ varies between 2/3
($p=5/3$) and 1/3 ($p=4/3$). In this region, where higher order FQH
fluids have been observed \cite {Pan2003}, the SF$^-$ mode occurs as
a well defined peak with energy independent of $B$. The higher order
FQH states that occur at $p-1=2/3$ and $p-1=1/3$ are the subject of
great current interest
\cite{Pan2003,Szlufarska2001b,Scarola2002,Chang2003}. It is possible
that the marked changes in lineshapes observed in the SF$^-$ spectra
shown in Fig. \ref{fig:roton} could be related to the appearance of
these higher order FQH liquids.
\par

A remarkably rich set of phenomena also exists beyond the primary
sequence of FQHE states in the lowest LL.  For $\nu < 1/3$,
additional FQHE sequences are found for $\nu = p / (\phi p \pm 1)$
with $\phi=4,6,8$ \cite{Pan2002}. There has been much recent interest
in the crossover region between different FQHE sequences, both near
$\nu=1/3$ \cite{Hirjibehedin2003a} and $1/5$
\cite{Jiang1990,Pan2002}. Beyond $1/5$, the system is thought to
condense into a Wigner solid \cite{Jiang1990}, which can undergo a
melting transition to FQHE states \cite{Pan2002}.
\par

Light scattering studies of quantum liquids with $\nu \leq 1/3$ is in
its infancy. Recent results display evidence of a surprising
coexistence between lower energy $\phi=4$ excitations and higher
energy $\phi=2$ modes \cite{Hirjibehedin2003a}. These experiments
also show that the crossover from $\phi=2$ to $\phi=4$ liquids that
happens at $\nu=1/3$ is associated with the remarkable emergence of
low lying excitations with energies $\omega < 0.2 \mathrm{meV}$.
\par

\section{FQHE States Around $\nu = 3/2$}
\label{sec:higherLL}

At lower $B$, FQHE states are found around $\nu=3/2$. These states
are also thought to have a CF-LL structure.  Evidence of crossings of
these levels is seen in angular-dependent magnetotransport
\cite{Clark1989,Eisenstein1989,Du1995,Du1997}. These states are
characterized by low lying excitations, making light scattering an
ideal probe of their structures.
\par

\begin{figure}
\includegraphics{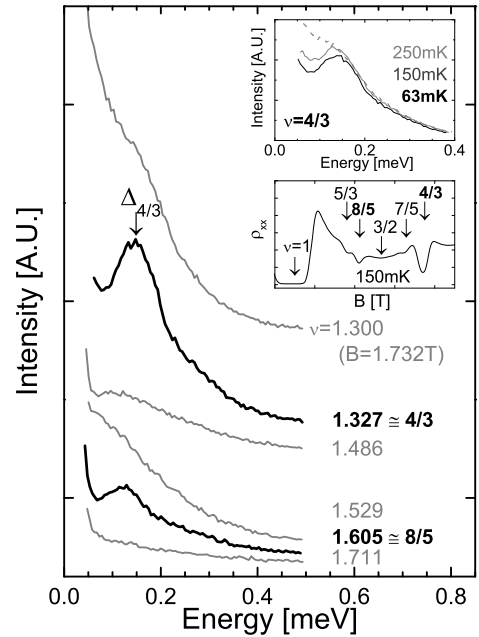}
\caption{\label{fig:higher_LL} Inelastic light scattering spectra at
various filling factors between $2 > \nu > 1$. The spectra at FQHE
states $\nu=4/3$ and $\nu=8/5$ are highlighted. The upper inset shows
the temperature dependence of the mode at $\nu=4/3$ and the lower
inset shows transport data identifying various FQHE states.}
\end{figure}

We have initiated light scattering studies of low lying excitations
of the 2D system in higher electron LLs with experiments carried out
in the filling factor range $2 > \nu > 1$.  Figure
\ref{fig:higher_LL} shows light scattering spectra from different
filling factors around $\nu = 3/2$. Strong excitations are seen at
FQHE states $4/3$ and $8/5$. Modes are seen only near these two FQHE
states and are not observed at other FQHE states that have been seen
in previous transport data
\cite{Clark1989,Eisenstein1989,Du1995,Du1997}, consistent with
magnetotransport shown in the lower inset of Fig.
\ref{fig:higher_LL}. These excitations are sharp ($<0.2\mathrm{meV}$
wide) and have resonance enhancement profiles that are relatively
narrow ($<0.4\mathrm{meV}$ wide, not shown). As seen in the upper
inset of Fig. \ref{fig:higher_LL}, the $4/3$ state has a very strong
temperature dependence. As the temperature is raised from $63
\mathrm{mK}$, the low energy tail increases in intensity until the
mode disappears above $300 \mathrm{mK}$.
\par

On the basis of particle-hole symmetry, the FQH states around $\nu =
3/2$ are described in terms of hole states in a filled spin-split LL
with hole filling factor $\nu_{H} = 2 - \nu$. Within the CF
framework, the CF quasiparticles fill quasi-LLs that are spaced by an
effective cyclotron energy $\omega_{CF}$ with $B^* = 3 (B -
B_{3/2})$, where $B_{3/2}$ is the value of $B$ at $\nu=3/2$, with
spin-split partners spaced by an effective Zeeman energy E$_z^* =g^*
\mu_B B_T$, where $g^*$ is an enhanced Lande factor including
many-body corrections to the spin flip energy \cite{Du1995}. The
angular dependent disappearance of minima at various FQHE states
\cite{Clark1989,Eisenstein1989} is interpreted as an energy level
crossing occurring when E$_z^* = j \omega_{CF}$ for integer $j$
\cite{Du1995,Du1997}.
\par

We focus our attention on $\nu = 4/3$, which being equivalent to a
$2/3$ state of holes has two CF-LLs populated. For small values of
$B_T$, the system is thought to be unpolarized, with the lowest
$|\uparrow \rangle$ and $|\downarrow \rangle$ states equally
populated. In a non-interacting picture, the lowest energy excitation
is from $|\downarrow \rangle \rightarrow | \uparrow \rangle$ and has
an energy $\Delta_{4/3} = \omega_{CF} - E_z^*$. Using this as the
form for the activation gap, Du et al. \cite{Du1997} find $m_{CF} =
0.43 m_0$ and $g^* \cong 0.76$, where $m_0$ is the free electron
mass. Assuming that $m_{CF}$ and $(g^*-g)$ scale with $e^2 / \epsilon
l_0$, we deduce $m_{CF} = 0.30 m_0$ and $g^* = 0.66$ for our sample.
Using these values, we find $\Delta_{4/3} = 0.15 \mathrm{meV}$ to be
in agreement with the peak in the excitation spectrum seen at $4/3$
in Fig. \ref{fig:higher_LL}.
\par

As seen in the upper inset to Fig. \ref{fig:higher_LL}, the mode at
$4/3$ disappears above $300 \mathrm{mK}$.  Similar behavior is also
seen for the mode at $8/5$.  The temperature dependence is much more
sensitive than the energy of $\Delta_{4/3}$.  The $4/3$ mode in Fig.
\ref{fig:higher_LL} also shows significant intensity below $0.1
\mathrm{meV}$, suggesting the existence of a lower energy roton. This
lower energy roton may determine the temperature dependence of the
state. Because of the relatively low density, it is difficult to
resolve the various features of the excitations described in Fig.
\ref{fig:disp}. Further work in higher density samples may allow us
to map the critical points in the dispersions of various modes of the
FQHE states around $\nu=3/2$ as has been done for the region between
$2/5 \geq \nu \geq 1/3$ \cite{Kang2001,Dujovne2003a,Dujovne2003b} and
$\nu \leq 1/3$ \cite{Hirjibehedin2003a}.
\par

It is surprising that clear excitation modes are not seen at the FQHE
state of $\nu = 5/3$. This is consistent with transport measurements
in our sample that indicate that  $5/3$ is not as well-defined as the
FQHE state at $8/5$, as seen in the lower inset of Fig.
\ref{fig:higher_LL}. Previous reports have indicated that the $5/3$
state is more robust than $8/5$
\cite{Willett1988,Clark1989,Eisenstein1989,Sajoto1990,Du1995}. Recent
results, however, have indicated that the $8/5$ state is stronger
than the $5/3$ state in very high quality samples at low densities
\cite{ZhuThesis}.
\par

\section{Summary}
\label{sec:summary}

Inelastic light scattering accesses the low lying quasiparticle
charge and spin excitations at and between fractional quantum Hall
states. Breakdown of wavevector conservation allows light scattering
to access energies of bound and unbound quasiparticle-quasihole pairs
and provides a measure of residual quasiparticle interactions. In the
filling factor range $2/5 \geq \nu \geq 1/3$ we find evidence that CF
quasiparticles have an electron-like LL scheme. Anomalies in the
magnetic field dependence of the SF excitations may be due to the
presence of higher order ($\phi=4$) CFs. In higher LLs, we find
excitations at $4/3$ and $8/5$ that are consistent with a
spin-unpolarized population of quasi-LLs. Future work will explore
higher order FQH states and the even denominator FQH states in higher
LLs.
\par

We wish to thank R. L. Willett for transport measurements. This work
was supported in part by the Nanoscale Science and Engineering
Initiative of the National Science Foundation under NSF Award Number
CHE-0117752 and by a research grant of the W. M. Keck Foundation.
\par



\end{document}